\documentclass[aps,prd,twocolumn,showpacs,superscriptaddress,groupedaddress]{revtex4}  
\usepackage{graphicx}  
\usepackage{dcolumn}   
\usepackage{bm}        
\usepackage{amssymb}   
\usepackage{multirow}
\usepackage[dvips]{color}

\begin{document}

\title{Nuclear effects in neutrino and antineutrino 
CCQE scattering at MINER$\nu$A kinematics}

\author{G.D.~Megias}
\affiliation{Departamento de F\'{\i}sica At\'{o}mica, Molecular y Nuclear, 
Universidad de Sevilla, 41080 Sevilla, Spain}

\author{M.V.~Ivanov}
\affiliation{Grupo\,de\,F\'{i}sica\,Nuclear,\,Departamento\,de\,
F\'{i}sica\,At\'omica,\,Molecular\,y\,Nuclear, 
Facultad\,de\,Ciencias\,\,F\'{i}sicas,\,Universidad\,Complutense\,de\,Madrid,
\,CEI Moncloa, Madrid\,E-28040,\,Spain}
\affiliation{Institute\,for\,Nuclear\,Research\,and\,Nuclear\,Energy,
\,Bulgarian\,Academy\,of\,Sciences,\,Sofia\,1784,\,Bulgaria}

\author{R.~Gonz\'alez-Jim\'enez}
\affiliation{Departamento de F\'{\i}sica At\'{o}mica, Molecular y Nuclear, 
Universidad de Sevilla, 41080 Sevilla, Spain}

\author{M.B.~Barbaro}
\affiliation{Dipartimento di Fisica, Universit\`a di Torino and  INFN, 
Sezione di Torino, Via P. Giuria 1, 10125 Torino, Italy}

\author{J.A.~Caballero}
\affiliation{Departamento de F\'{\i}sica At\'{o}mica, Molecular y Nuclear, 
Universidad de Sevilla, 41080 Sevilla, Spain}

\author{T.W.~Donnelly}
\affiliation{Center for Theoretical Physics, Laboratory for 
Nuclear Science and Department of Physics,
Massachusetts Institute of Technology, Cambridge, Massachusetts 02139, USA}

\author{J.M.~Ud\'ias} 
\affiliation{Grupo\,de\,F\'{i}sica\,Nuclear,\,Departamento\,de\,
F\'{i}sica\,At\'omica,\,Molecular\,y\,Nuclear, 
Facultad\,de\,Ciencias\,\,F\'{i}sicas,\,Universidad\,Complutense\,de\,Madrid,
\,CEI Moncloa, Madrid\,E-28040,\,Spain}


\begin{abstract}
We compare the charged-current quasielastic neutrino and
antineutrino observables obtained in two different nuclear models,
the phenomenological SuperScaling Approximation and the Relativistic
Mean Field approach, with the recent data published by the
MINER$\nu$A Collaboration. Both models provide a good description of
the data without the need of an {\it ad hoc} increase in the mass
parameter in the axial-vector dipole form factor. Comparisons are
also made with the MiniBooNE results where different conclusions are
reached.

\end{abstract}

\maketitle

\section{Introduction}

The MINER$\nu$A Collaboration has recently measured differential
cross sections for neutrino and antineutrino charged-current
quasielastic (CCQE) scattering on a hydrocarbon 
target~\cite{Fiorentini:2013ezn,Fields:2013zhk}. 
``Quasielastic'' events
are defined, in this case, as containing no mesons in the final
state. The beam energy goes from $1.5$ to $10$~GeV and is peaked at
$E_{\nu}\sim 3$~GeV. At lower energies $E_\nu\sim 0.7$~GeV the
MiniBooNE experiment has
reported~\cite{AguilarArevalo:2010zc,AguilarArevalo:2013hm} CCQE
cross sections that are higher than most theoretical predictions based on the
impulse approximation (IA), leading to the suggestion that non-QE
processes induced by two-body currents may play a significant role
in this energy
domain~\cite{Martini:2010ex,Amaro:2010sd,Amaro:2011aa,Nieves:2011yp}.
These effects have sometimes been simulated, in the Relativistic
Fermi Gas (RFG) framework, by a value of the nucleon axial-vector
dipole mass
$M_A=1.35$~GeV~\cite{AguilarArevalo:2010zc,AguilarArevalo:2013hm},
which is significantly larger than the standard value
$M_A=1.032$~GeV extracted from neutrino-deuterium quasielastic
scattering. On the other hand, higher-energy data from the NOMAD
experiment ($E_\nu\sim 3-100$~GeV) are well accounted for by IA
models~\cite{Amaro:2013yna}. The MINER$\nu$A experiment is situated
in between these two energy regions and its interpretation can
therefore provide valuable information on the longstanding problem
of assessing the role of correlations and meson exchange currents
(MEC) in the nuclear dynamics~\cite{Donnelly:1978xa,De
Pace:2003xu,Amaro:2010iu}.

In this paper we present results corresponding to two different
nuclear models: the SuSA (SuperScaling Approximation) and the RMF
(Relativistic Mean Field) approach. Both have been extensively
tested against existing QE electron scattering data over a wide
energy range. The detailed description of these models can be found in
our previous work (see, {\it e.g.}, \cite{Amaro:2004bs} and
\cite{Amaro:2011qb}). Here we just summarize their main features and
address some improvements with respect to previous work.

\section{Results}

SuSA~\cite{Amaro:2004bs} is based on the idea of using electron
scattering data to predict CC neutrino cross sections: a
phenomenological ``superscaling function'' $f(\psi)$, depending only
on one ``scaling variable'' $\psi(q,\omega)$  and embodying the
essential nuclear dynamics, can be extracted from QE longitudinal
$(e,e')$ data within a fully relativistic framework. This function
is then multiplied by the appropriate charge-changing N$\to$N 
($n\to p$ for neutrino and $p\to n$ for antineutrino scattering) weak
interaction cross sections to obtain the various response functions
that contribute to the inclusive neutrino-nucleus cross section. On
the one hand, the model gives a good representation of the purely
nucleonic contributions to the existing QE electron scattering data,
to the extent that quasielastic scattering
can be isolated. On the other hand, it does not account for
inelastic scattering and MEC which are mainly
seen in the transverse channel. For the former, the SuSA approach
has been successfully extended to higher energies into the non-QE
regime where inelastic contributions dominate~\cite{Maieron:2009an}.
The latter have been modeled using extensions of the RFG for
two-body operators and typically cause $10-20$\% scaling
violations. 

The model works well for high enough momentum and energy transfers,
whereas in the low $q$ and $\omega$ region (typically,
$q\le400$~MeV/c and $\omega\le50$~MeV) it is inadequate and
different approaches which account for Pauli blocking and collective
nuclear excitations should be used. In the phenomenological SuSA approach, Pauli blocking effects are not trivial to implement and have been
neglected so far in our previous applications of the model. In this work
we introduce them using the procedure proposed in
\cite{Rosenfelder:1980nd}, which generalizes the simple RFG prescription -- only valid for a step-like momentum distribution -- to accommodate more realistic momentum distributions. In summary, the prescription consists in
subtracting from the scaling function $f(\psi(\omega,q))$ its mirror
function $f(\psi(-\omega,q))$: this, as argued in that reference,
incorporates a correct blocking of unphysical excitations, which are then excluded in a more satisfactory way than through the {\it ad hoc} factor $[1-n(\mathbf{p}+\mathbf{q})/n(0)]$ commonly used in the literature. 
If applied to a
non-Pauli-blocked version of the RFG, this procedure yields exactly
the correct Pauli blocking for that model. 
Moreover, this method does not require the knowledge of the nucleon
momentum distribution $n(p)$. Additionally, Coulomb
corrections for the outgoing lepton are taken into account in the
SuSA approach~\cite{Amaro:2004bs} and a phenomenological energy
shift $E_{\text{shift}}=20$~MeV is introduced in the scaling
variable in order to reproduce the correct peak position of electron
scattering data~\cite{Chiara}.

\begin{figure}[htcb]
\includegraphics[scale=0.26, bb=22 12 970 724, clip]{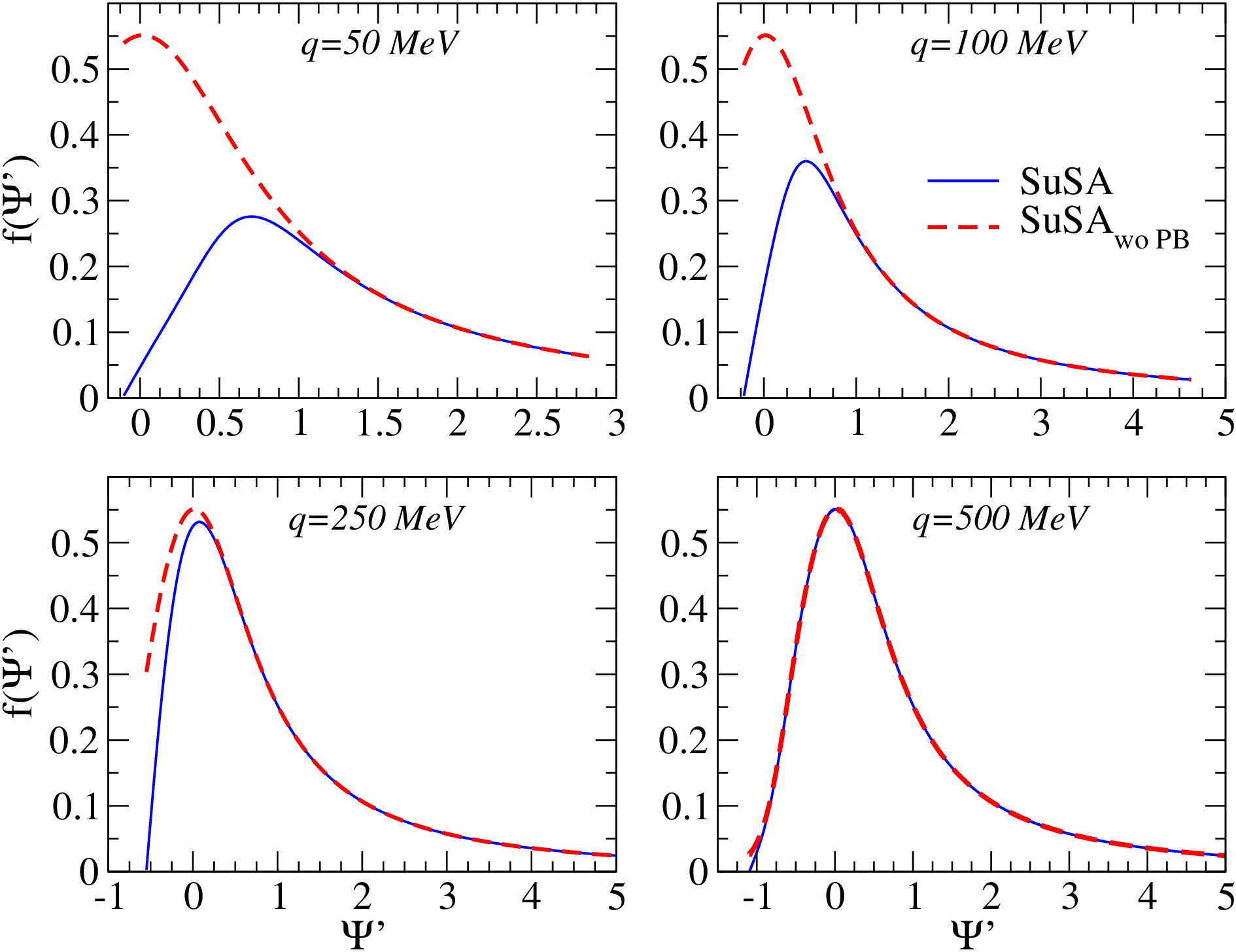}\\%
\caption{(Color online) Superscaling function versus $\psi'$ at different q-fixed values and evaluated for the SuSA model with ($SuSA$) and without ($SuSA_{woPB}$) Pauli blocking.
.\label{fig:figintro1}}
\end{figure}

In  Fig.~\ref{fig:figintro1},  Pauli blocking effects in the scaling function for different fixed values of the transferred 
momentum are shown. Pauli blocking effects are  noticeable when $q<250$ MeV and the subtractions to the non-Pauli-blocked scaling function come mainly from $\omega<50$ MeV.

The second model we consider is the RMF, where the nucleons' wave
functions are, for both the bound and scattering states, solutions
of the Dirac-Hartree equation in the presence of strong,
energy-independent, real scalar attractive and vector 
repulsive potentials. The model fulfills dispersion relations and
maintains the continuity equation~\cite{Horikawa}. In the RMF model
the nucleons are dynamically and strongly {\em off-shell} and,
 as a consequence the cross section is not factorized into a
spectral function and an elementary lepton-nucleus cross section, as
happens in other approaches~\cite{Benhar:2010nx}. In order to
appreciate the effects of off-shellness on the RMF results we will
also show results in which the spinors are put exactly on the mass
shell, within the so-called effective momentum approach
(EMA)~\cite{Udias:1993xy}.

\begin{figure}[htcb]
\includegraphics[scale=0.25, bb=-12 22 970 647, clip]{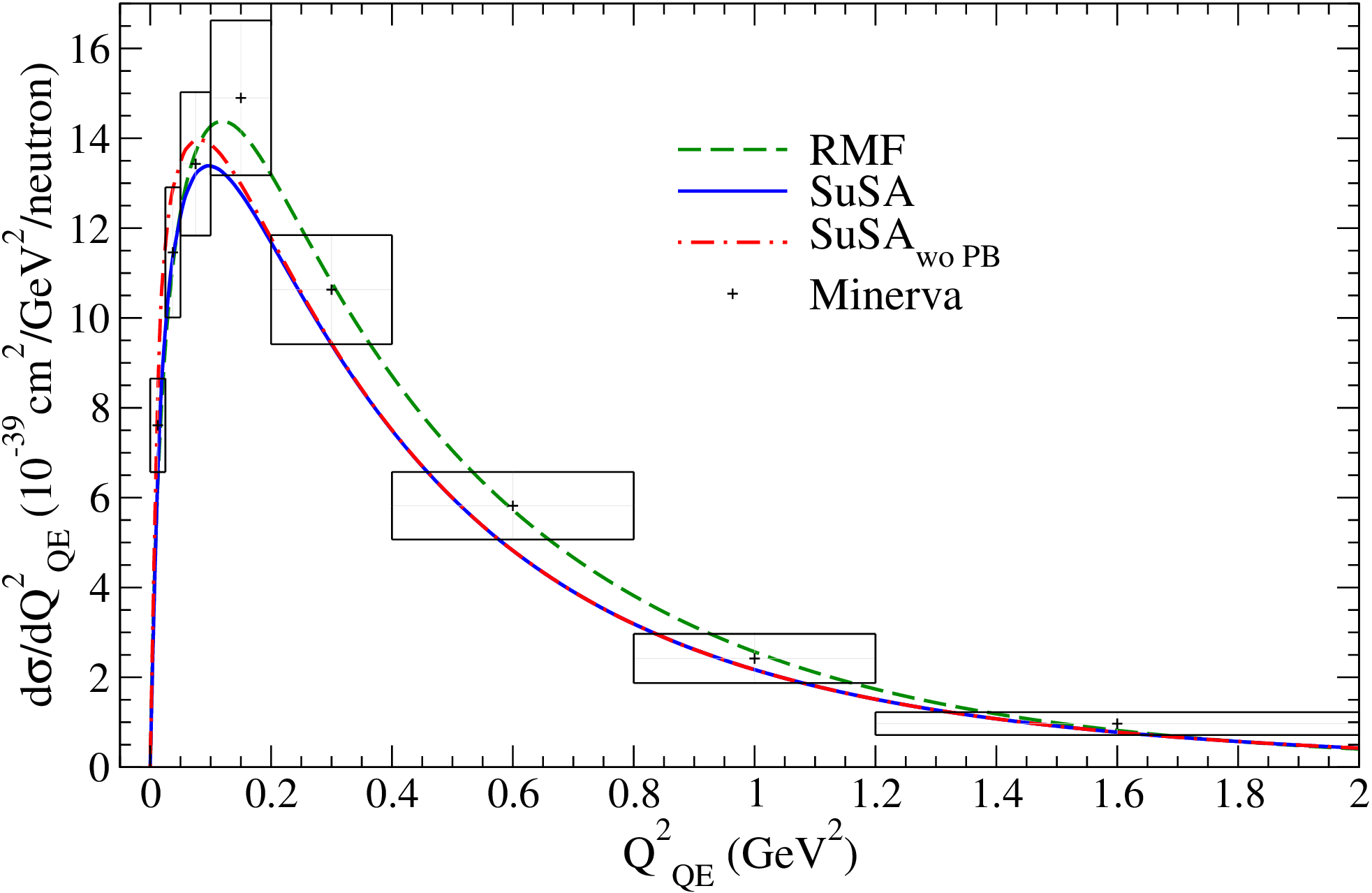}\\%
\caption{(Color online) Unfolded CCQE $\nu_\mu-^{12}$C scattering cross section per target
nucleon as a function of $Q^2_{QE}$ at fixed neutrino energy $E_\nu$=3 GeV and evaluated in the RMF and in the SuSA (with and without Pauli-blocking) models. MINER$\nu$A data are from~\cite{Fiorentini:2013ezn,Fields:2013zhk}
.\label{fig:figintro2}}
\end{figure}

Before entering into the comparison of fully folded results with the neutrino spectrum results, first in~Fig.~\ref{fig:figintro2} 
the unfolded CCQE neutrino cross section at Minerva kinematics for a fixed neutrino energy  of 3 GeV is presented,  evaluated within the RMF and SuSA models, with and without Pauli blocking.  It can be seen that Pauli blocking softly decreases the cross section 
at low $Q^2_{QE}$, which is directly related to the higher contribution of the  low $q$ and $\omega$ kinematic region in this case. 
Note also that our theoretical results for a fixed $E_\nu$ value (near the peak of the flux) are in good agreement with 
the MINER$\nu$A data, as also observed in \cite{Nieves:2013min}.
It is interesting to see that, for $|Q^2_{QE}|>0.2$ (GeV/c)$^2$, the RMF cross section is slightly higher than the SuSA results. 

\begin{figure}[htcb]
\includegraphics[scale=0.25, bb=-12 60 970 647, clip]{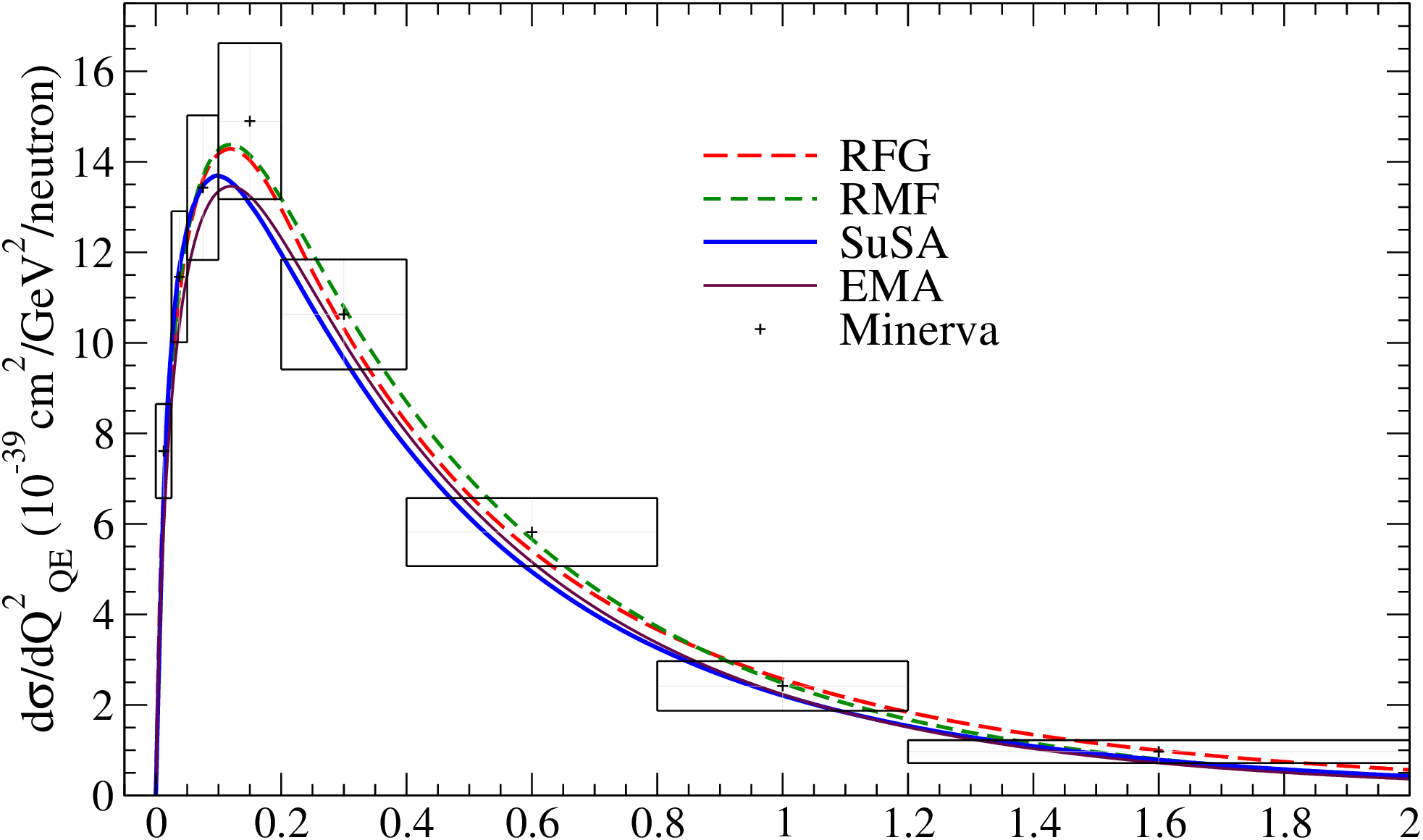}\\%
\includegraphics[scale=0.25, bb=-12 22 970 647, clip]{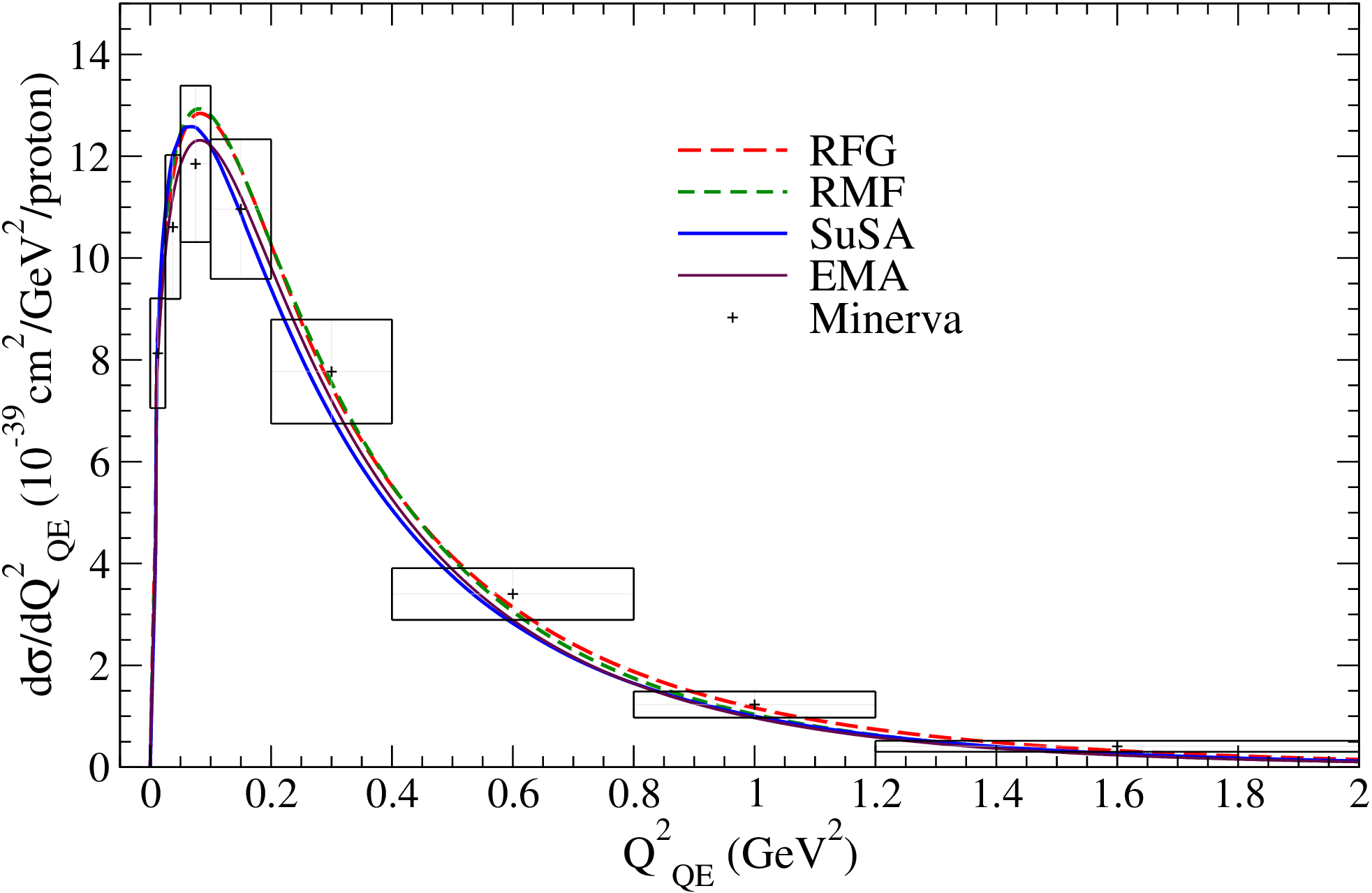}
\caption{(Color online) Flux-folded CCQE $\nu_\mu-^{12}$C (upper panel) and 
$\overline\nu_\mu-$CH (lower panel) scattering cross section per target
nucleon as a function of $Q^2_{QE}$ and evaluated in the SuSA, RMF and 
EMA models. 
MINER$\nu$A data are from~\cite{Fiorentini:2013ezn,Fields:2013zhk}
.\label{fig:fig1}}
\vspace{-0.5cm}
\end{figure}

In Fig.~\ref{fig:fig1} we display the flux-folded differential cross section
$\displaystyle{d\sigma/dQ^2_{QE}}$ for both neutrino (upper panel)
and antineutrino (lower panel) scattering off a CH target as a
function of the reconstructed four-momentum transfer squared
($Q^2_{QE}$), that is obtained in the same way as for the
experiment, assuming an initial-state nucleon at rest with a constant 
binding energy, $E_b$, set to $34$~MeV ($30$~MeV) in the neutrino
(antineutrino) case.
The cross sections are folded with the MINER$\nu$A $\nu_\mu$ and 
$\overline\nu_\mu$ fluxes~\cite{Fiorentini:2013ezn,Fields:2013zhk}
and the nucleon's axial mass has the standard value $M_A=1.032$~GeV.
We observe that both SuSA and RMF models yield predictions in
excellent agreement with the experimental data, leaving not much
space for large effects of 2p2h contributions, although perhaps $\sim$10$\%$ 
additional effects from MEC are acceptable. RMF results are
slightly higher than the SuSA ones, an outcome already observed at
MiniBooNE kinematics (see also Fig.~\ref{fig:figintro2}), which is related to the lower component
enhancement of the RMF spinors. Indeed, the EMA curves, where such
off-shell effects are absent,  lie closer to the SuSA results. 
The RFG model is also shown for reference. In the RFG calculation we use the formalism of \cite{Alberico:1988bv}, assuming a Fermi momentum of 228 MeV/c and an energy shift of 20 MeV. This is not the same as the RFG modeling of GENIE~\cite{GENIE} and NuWRO~\cite{NuWRO}, which could explain the slight difference between our RFG results and the ones reported in \cite{Fiorentini:2013ezn,Fields:2013zhk}. Note
that the RFG model with the standard value of the axial mass 
(red-dashed curve) also fits the data, being in very good agreement with the
other approaches, in particular with RMF. Finally, the spread in the
curves corresponding to the four models is less than $7$\% 
in the case of neutrinos and less than
$5$\% in the case of antineutrinos (see discussion below). 
The theoretical results presented here include the whole energy range for the neutrino.
The experimentalists implement several cuts on the phase space of the data, such as
restricting the kinematics to contributions from neutrino energies below 10 GeV.   
The impact of such a cut on the results we present here is smaller than 0.2$\%$, 
in the worst case.
In the experimental analysis, several cuts were imposed to the initial data sample
to increase the ratio of true quasielastic events in the sample. The effect of 
these cuts has been incorporated into the efficiency factors of 
the experiment and thus the data have been corrected for them~\cite{expthesis:2012}. 
We apply no cuts to the theoretical results, as the data have been corrected for their effect.

\begin{figure}[t]
\includegraphics[scale=0.25, bb=-12 60 970 647, clip]{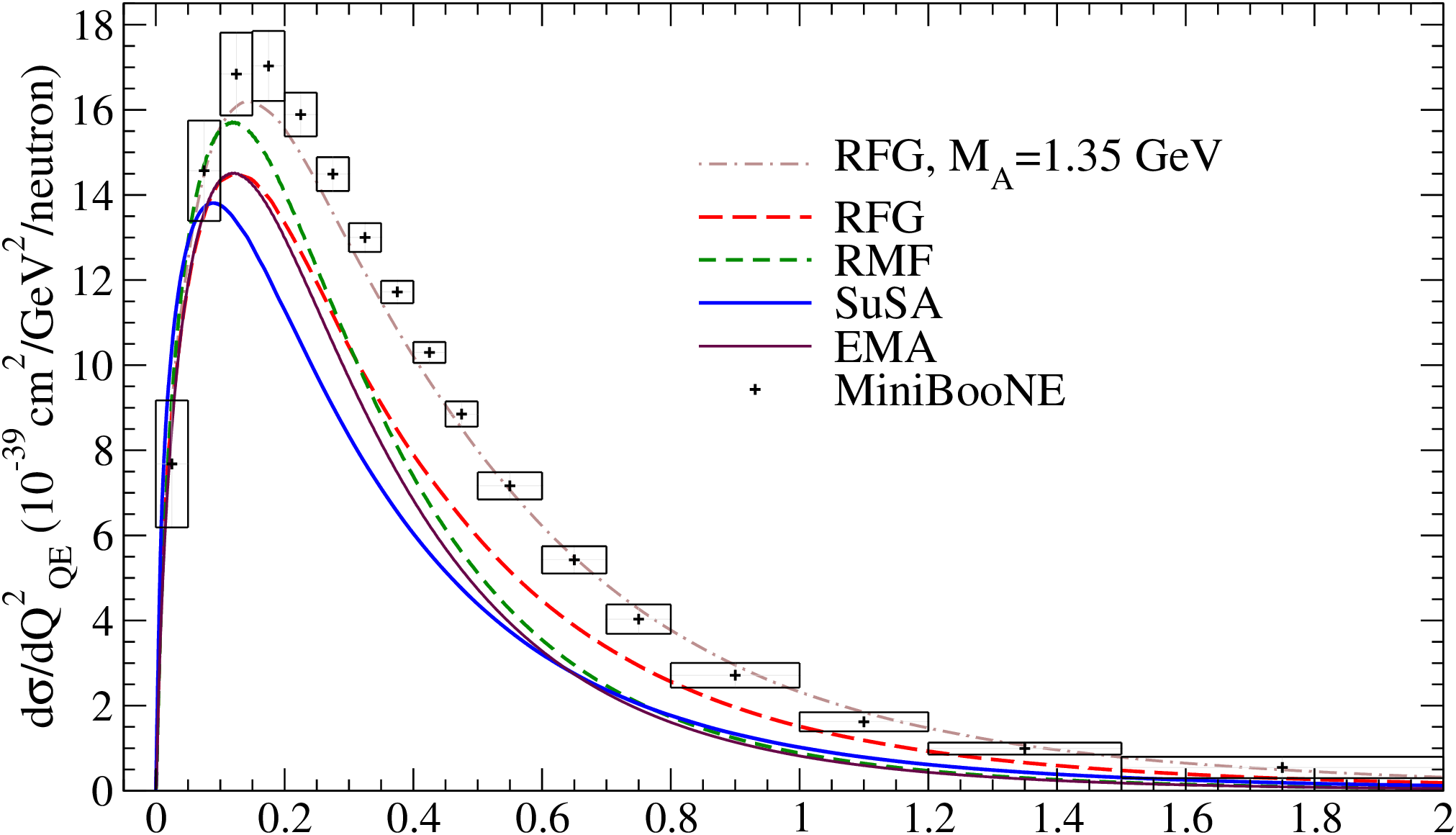}\\%
\includegraphics[scale=0.25, bb=-12 22 970 647, clip]{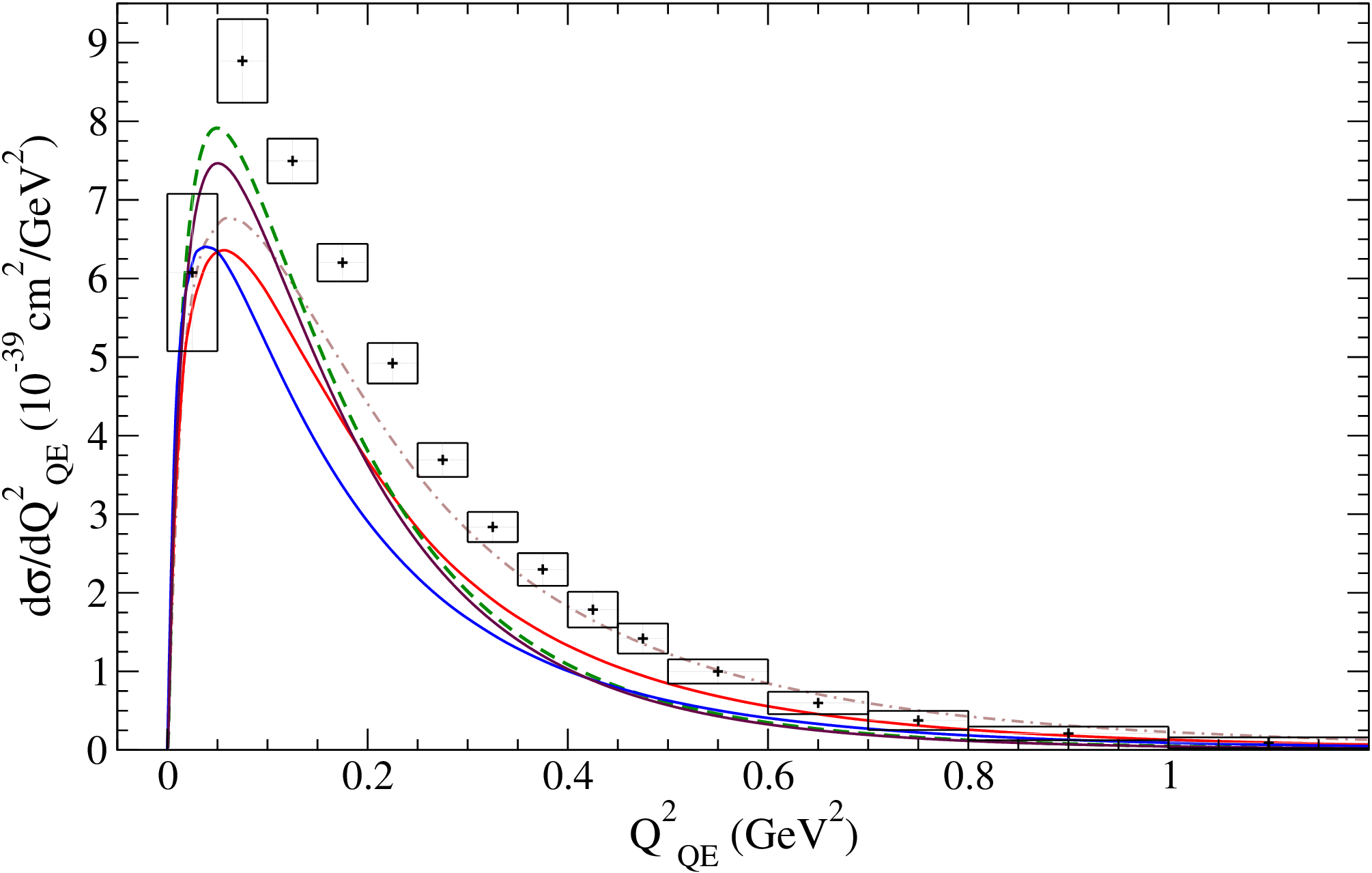}
\caption{(Color online) Flux-folded CCQE $\nu_\mu-^{12}$C (upper panel) and 
$\overline\nu_\mu-^{12}$C (lower panel) scattering cross section per target
nucleon as a function of $Q^2_{QE}$ and evaluated in the SuSA, 
RMF and EMA models and compared with MiniBooNE
data~\cite{AguilarArevalo:2010zc,AguilarArevalo:2013hm}. 
The RFG model is shown for two values of the axial mass (see text for details).
\label{fig:fig2}}
\vspace*{-0.18cm}
\end{figure}

For completeness we illustrate in Fig.~\ref{fig:fig2} the 
differential cross section $d\sigma/dQ^2_{QE}$ corresponding to the
MiniBooNE experiment. The same qualitative behaviour among the
models is observed here as for MINER$\nu$A kinematics. Namely, the SuSA
approach provides the lowest cross section and RFG/RMF the highest
one, and as already shown in the previous figure, the EMA curves
come closer to SuSA. However, the spread among the different
theoretical predictions is larger for MiniBooNE, about twice as much
as for MINER$\nu$A. 
Further, in contrast to the MINER$\nu$A experiment,
all models exhibit a different energy dependence and underestimate 
the MiniBooNE data, unless the axial mass in the dipole
parameterization of the axial-vector form factor is significantly
increased (see the RFG curve for $M_A=1.35$~GeV). Note also that the
MiniBooNE Collaboration reproduces their $d\sigma/dQ^2_{QE}$
measurements by normalizing their RFG predictions ($M_A=1.35$~GeV) 
to the observed total cross section. Moreover, although MiniBooNE data
errorbands are much smaller than the ones corresponding to
MINER$\nu$A, the comparison between theory and data shows a clear
difference between the two situations: whereas in the former
(MiniBooNE) no model based on the IA is capable of reproducing the
data, in the latter (MINER$\nu$A) the IA already provides a good
description of data and the enlargement of the axial mass worsens
this agreement.

\begin{figure}[htcb]
\includegraphics[scale=0.25, bb=-12 22 970 647, clip]{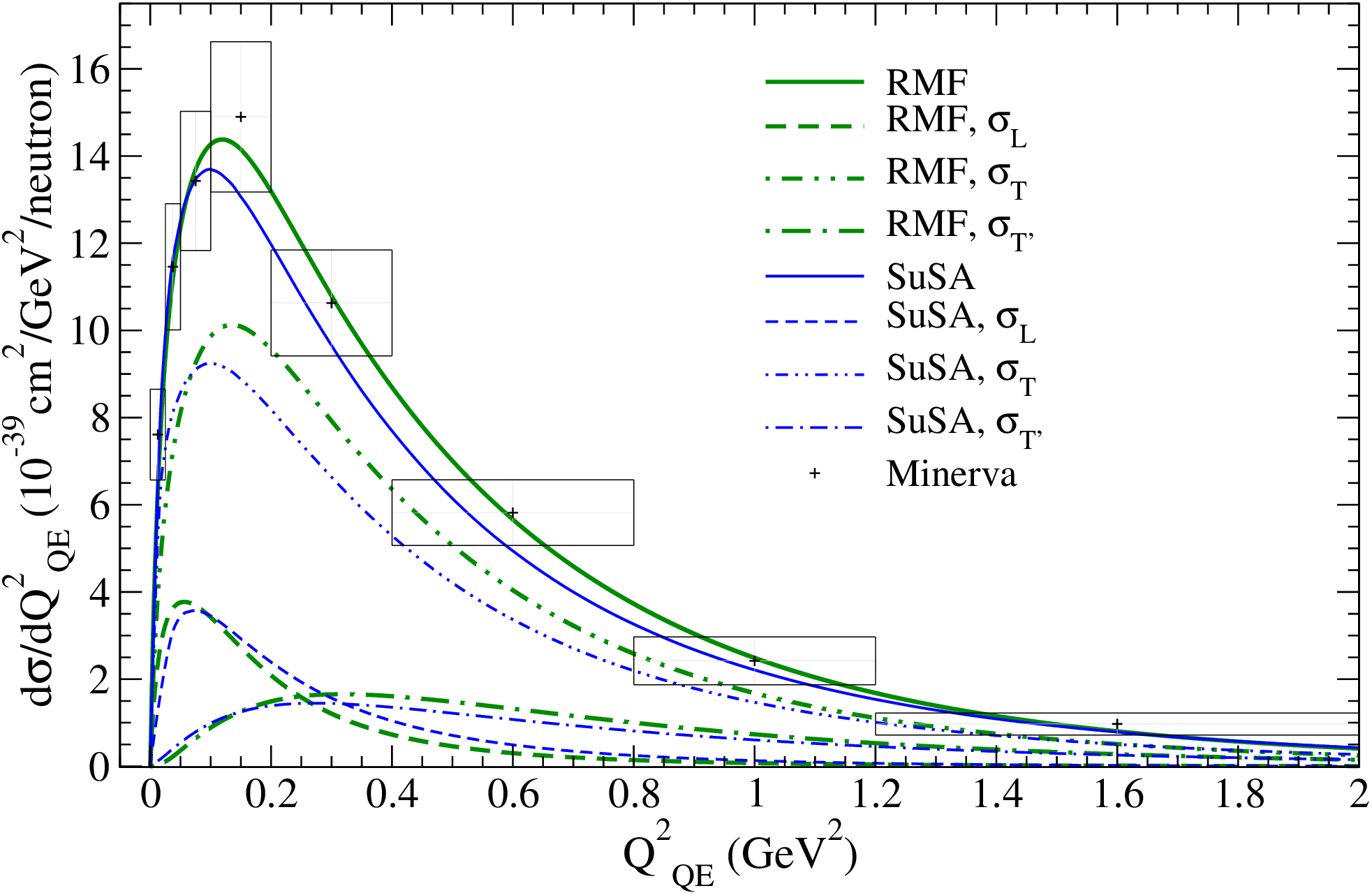} 
\caption{(Color online) Separated contributions of CCQE $\nu_\mu-^{12}$C $d\sigma/dQ^2_{QE}$ 
in the SuSA and RMF models.}\label{fig:fig3}
\end{figure}

In Fig.~\ref{fig:fig3} we display the separate contributions of the
longitudinal $(L)$, transverse $(T)$ and transverse-axial
interference $(T')$ channels to the differential cross section
within the two models, SuSA and RMF, showing that the transverse
response is dominant in the full range of $Q^2_{QE}$. As observed,
the difference between SuSA and RMF results is mostly linked to the
$T$ response. Moreover, the different role played by the
interference $T'$ response for neutrinos (constructive) and
antineutrinos (destructive) explains the overall difference between
SuSA and RMF curves for the cross section, being larger for
neutrinos (Figs.~\ref{fig:fig1} and~\ref{fig:fig2}).

\begin{figure}[htcb]
\includegraphics[scale=0.25, bb=-12 60 970 647, clip]{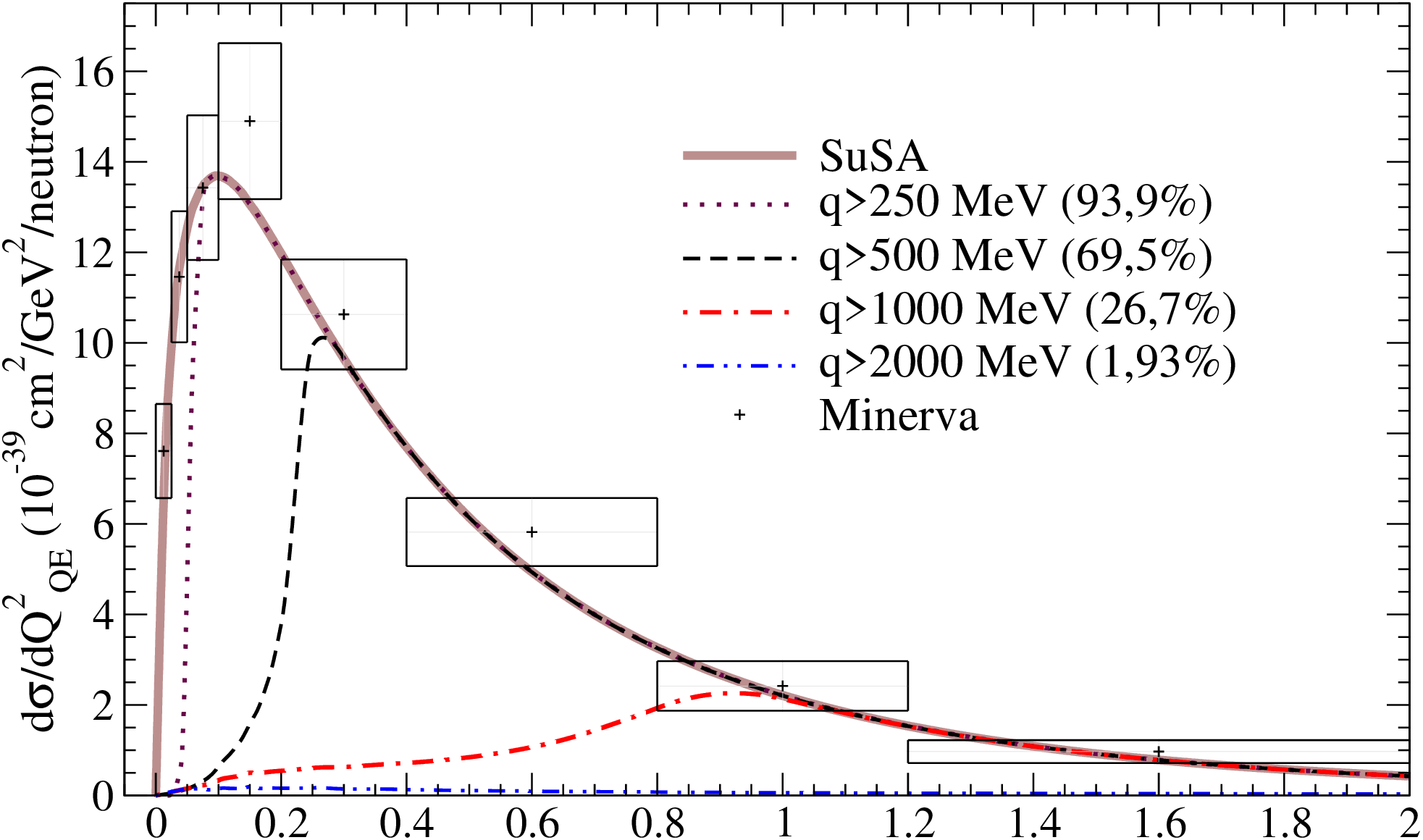}\\%
\includegraphics[scale=0.25, bb=-12 22 970 647, clip]{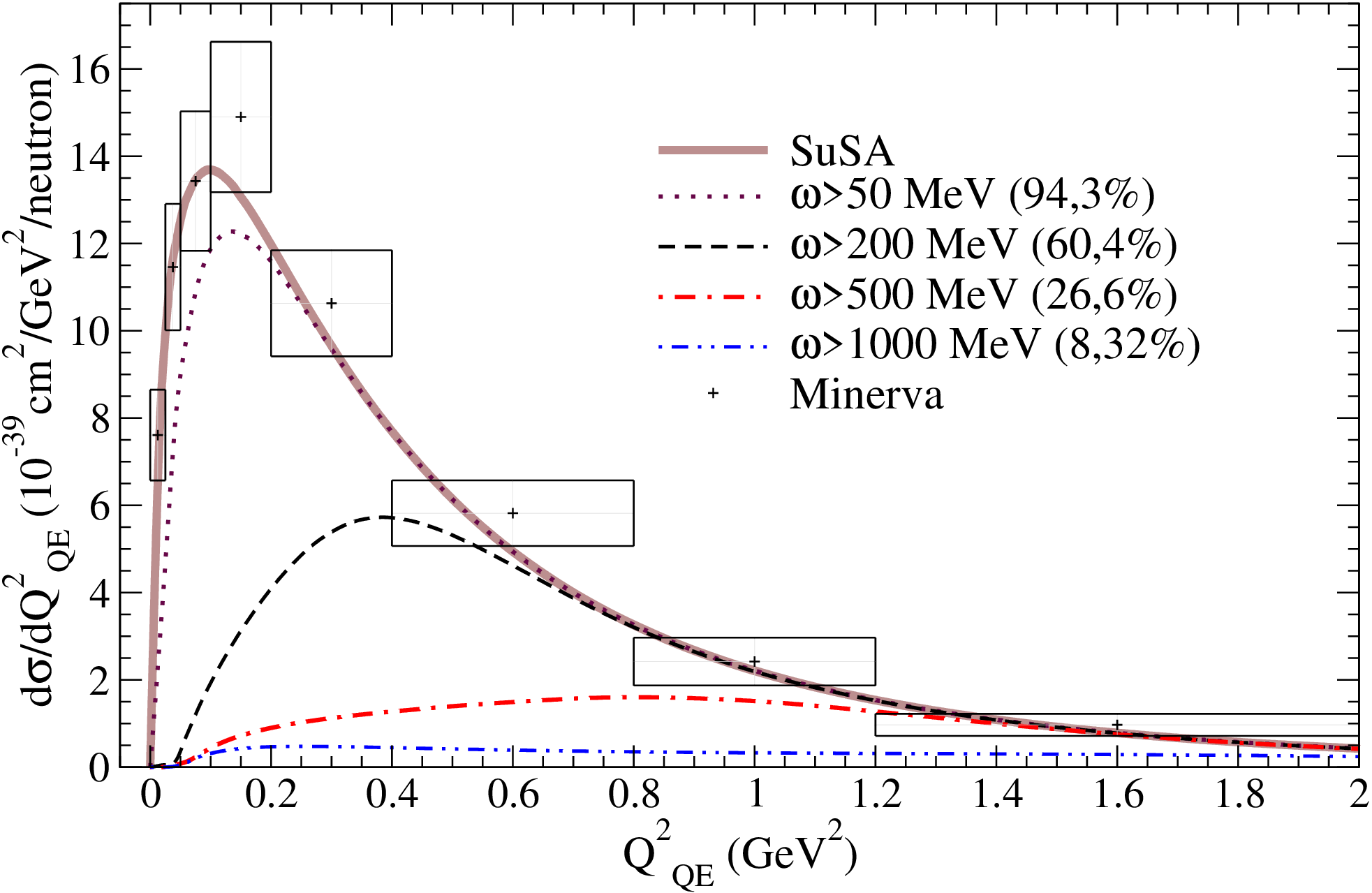}
\caption{(Color online) Flux-folded CCQE $\nu_\mu-^{12}$C differential cross section per 
target nucleon evaluated excluding all contributions
coming from transferred momentum (upper panel) and energy (lower panel) 
below some selected values, as indicated in the figure.
MINER$\nu$A data are from~\cite{Fiorentini:2013ezn,Fields:2013zhk}. 
Numbers in parentheses refer to the fraction of the total cross
section corresponding to each curve.\label{fig:fig4}}
\vspace{-0.3cm}
\end{figure}

It was shown in \cite{Amaro:2013yna} that, even at {\it high}
neutrino energies, {\it low} energy and momentum transfers play a
crucial role in the CCQE cross section. To illustrate this point in
the specific conditions of MINER$\nu$A, we display in
Fig.~\ref{fig:fig4} the neutrino cross section evaluated in the SuSA
model by applying different cuts in $q$ (upper panel) and $\omega$
(lower panel): it clearly appears that, even if the neutrino energy
is as large as $3$~GeV, the process is largely dominated by
energy and momentum transfer, namely, $\omega<50$ MeV, $q<1000$ MeV. 
In Fig.~\ref{fig:fig4} we also give
the relative contribution to the cross section (expressed in
percentage) attached to the different $(q,\omega)$ regions
considered. Note how the relative fraction diminishes very
significantly for increasing $q$, $\omega$ values.

In Table~\ref{tab:table} we report the values of the total cross
sections per nucleon integrated over the flux from $1.5$ to
$10$~GeV, for both neutrino and antineutrino scattering: the results
corresponding to all models (RFG, SuSA, RMF and EMA) are compatible
with the experimental data within the errorbars. The discrepancy
between theory and data (central values) is at most of the order 
of $\sim 9-10\%$ (SuSA/EMA), being reduced to $\sim 2-3\%$ for RMF/RFG.
\begin{table}[htcb]
\vspace{-0.05cm}
\caption{\label{tab:table} (Color online) Comparisons between the measured total
cross section (per nucleon) after averaging over the flux and the
results obtained with the RFG, SuSA and RMF models.}
\begin{ruledtabular}
\begin{tabular}{c|ccccc}
Model&RFG&\mbox{SuSA}&\mbox{RMF}&\mbox{EMA}&Experimental\\
\hline
$\sigma_{\nu_\mu} (10^{-38}\mbox{cm}^2)$&0.916&0.834&0.901& 0.828 
&0.93$\pm$0.12\\
\hline
$\sigma_{\bar\nu_\mu} (10^{-38}\mbox{cm}^2)$&0.601& 0.550 & 0.583 
&0.554&0.604$\pm$0.083  \\
\end{tabular}
\end{ruledtabular}
\vspace*{-0.1cm}
\end{table}

\begin{figure}[htcb]
\includegraphics[scale=0.25, bb=-12 60 970 647, clip]{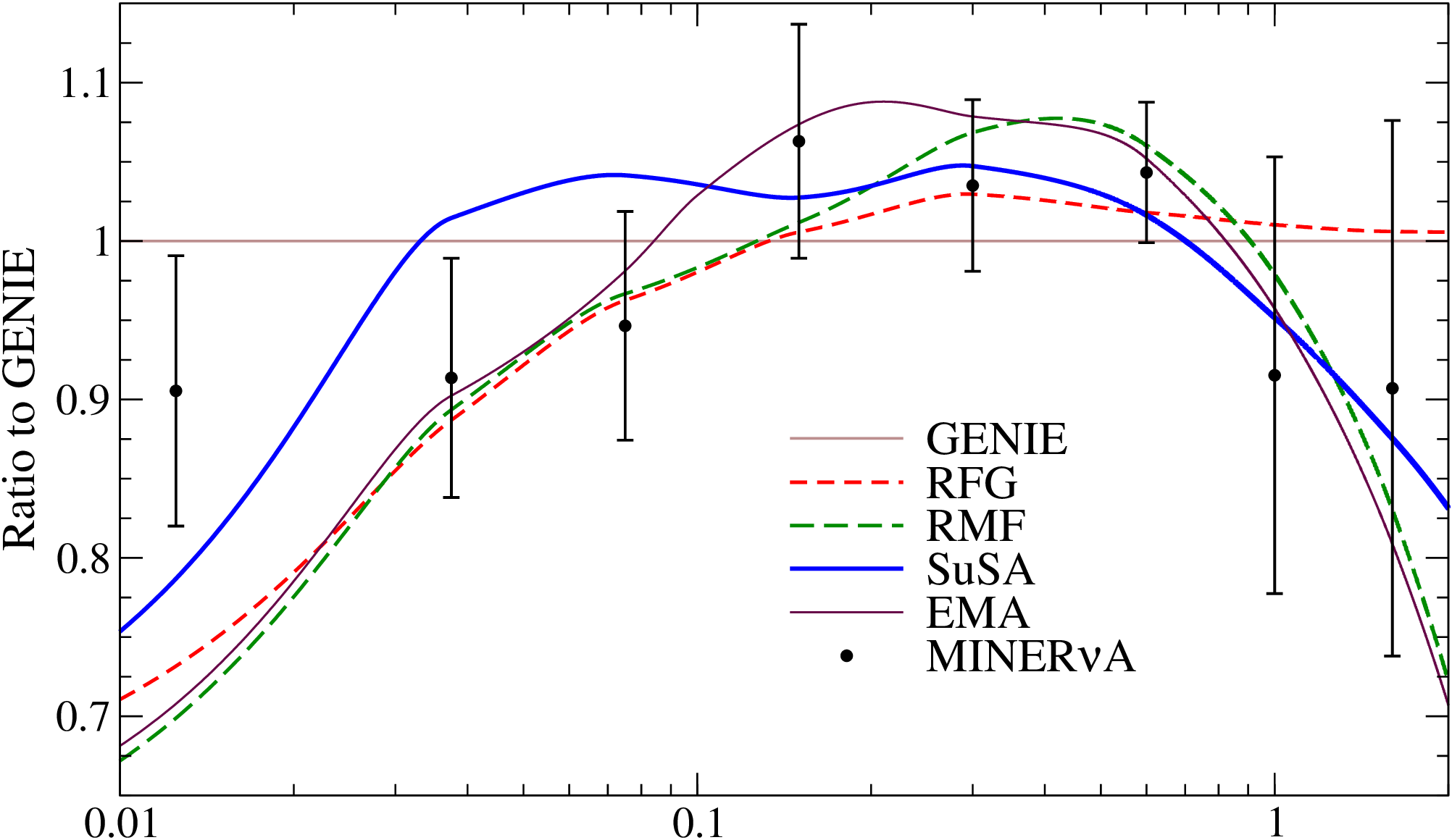}\\  
\includegraphics[scale=0.25, bb=-12 22 970 647, clip]{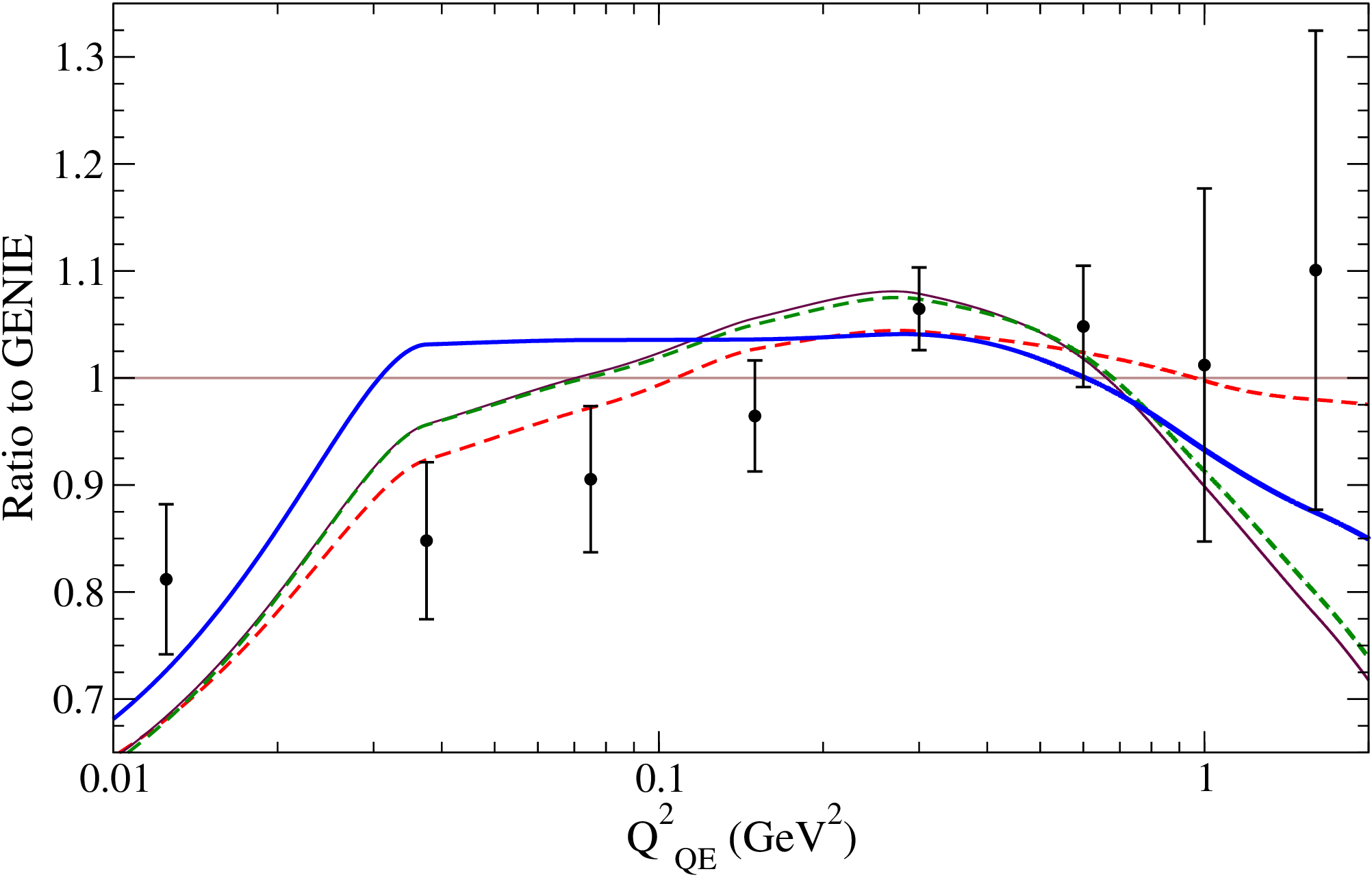}
\caption{(Color online) The MINER$\nu$A data and models of Fig.~\ref{fig:fig1} shown versus 
$Q^2_{QE}$ as a ratio to the GENIE prediction. Upper panel: neutrino case. 
Lower panel: antineutrino case.\label{fig:fig5}}
\vspace*{0.15cm}
\end{figure}

Finally, in Fig.~\ref{fig:fig5} the data and models are shown versus
$Q^2_{QE}$ as a ratio to the GENIE~\cite{GENIE} prediction, in the same way they
are presented by the MINER$\nu$A
Collaboration~\cite{Fiorentini:2013ezn,Fields:2013zhk}. The ratio
has the advantage of minimizing systematic uncertainties and better
emphasizing the differences between various models. More
specifically, the results are obtained by dividing each theoretical
model and the experimental data by the GENIE result and normalizing
these results to have the same total cross section across the range
$Q^2\in[0,2]$~GeV$^2$ as GENIE has. 
As shown, all the theoretical results, except SuSA, are within the error
bars of all but the lowest $Q^2$ data for the neutrino ratio,
while there is a slight overestimation in the central $Q^2$ data for the
antineutrino case. 
Actually the SuSA curve departs from the other models for the lower three data points. 
Note however that the ratio is strongly affected by the above mentioned normalization. Morever, this is the region where
the differential cross section reaches its maximum and changes its
shape dramatically, which, in addition to the reduced size of the
bins in this region, makes it difficult to compare accurately theory
and data. For higher $Q^2_{QE}$ the agreement of theory and data
improves for neutrinos (upper panel) where all theoretical results
lie within the data errorbands except SuSA for $Q^2_{QE}$-bins in
the range $[0.025,\, 0.1]$~GeV$^2$. For antineutrinos theory lies
above data for  $Q^2\in[0.025,0.2]$~GeV$^2$. In all the cases,
neutrinos and antineutrinos, the differences between theoretical
predictions are larger at the extreme $Q^2_{QE}$-bins, being
significantly reduced within the central values of $Q^2_{QE}$ where
the comparison with data is also much better.

\begin{table}[htcb]
\caption{\label{tab:tablexi2} (Color online) Comparisons between the measured
$d\sigma/dQ^2_{QE}$ (and its shape in $Q^2_{QE}$) and model
predictions, expressed as $\chi^2$ per degree of freedom (d.o.f) for
eight (seven) degrees of freedom.}
\begin{ruledtabular}
\begin{tabular}{c|c|cccc}
 &Model&RFG&\mbox{SuSA}&\mbox{RMF}&\mbox{EMA}\\
\hline
\hline
\multirow{2}{*}{$\nu_\mu$} &Rate $\chi^2/$d.o.f & 1.62 & 2.98 & 2.58 & 2.19 \\
 &Shape $\chi^2/$d.o.f & 1.82 & 4.00 & 2.90 & 2.87 \\
 \hline
\multirow{2}{*}{$\bar\nu_\mu$} &Rate $\chi^2/$d.o.f & 3.23 & 3.59 
& 3.92 & 3.52 \\
 &Shape $\chi^2/$d.o.f & 3.69 & 4.88 & 4.66 & 4.65 \\
\end{tabular}
\end{ruledtabular}
\end{table}

In Table~\ref{tab:tablexi2} we present the results obtained through
a $\chi^2$ test using cross sections (rate) and fractions of cross
sections (shape) for neutrinos and antineutrinos, and considering
the four models: SuSA, RMF, EMA and RFG. This test allows us to
estimate quantitatively the level of agreement between data and
predictions, accounting for the significant correlations between the
data points. Note that the fit analysis seems to work better for
neutrinos and the $\chi^2$-values are slightly smaller in the case
of the {\sl ``rate''} observable. The values obtained for $\chi^2$
indicate that there are some differences between all of the
theoretical models and the data. As seen in Fig.~\ref{fig:fig5}, all
of the models considered in this work fall below GENIE's predictions
for the larger and smaller bins in $Q^2_{QE}$. Where experimental
uncertainties are small enough to draw conclusions the same trend
appears to be seen in the data. 
Although not shown we have checked that the $\chi^2$-fit improves very 
significantly if the lowest $Q^2_{QE}$-value is removed from the analysis.

\section{Conclusions}

Summarizing, we have presented predictions for the differential
cross sections corresponding to the MINER$\nu$A experiment with two
nuclear models, SuSA and RMF. Both models are based on the IA and
work nicely in describing QE $(e,e')$ data. Contrary to previous
studies for the MiniBooNE experiment, we have shown that the two
models provide a good description of MINER$\nu$A data without the
need of increasing the nucleon axial mass and without having to
invoke any significant contributions from 2p2h MEC. Finally, a
discussion of results for the ratios to GENIE has been also
presented.

Our present studies, in addition to previous ones applied to the
MiniBooNE and NOMAD experiments, seem to indicate either some
inconsistency between these experiments (for example in the
definition of what is ``quasielastic'' and what is ``pion
production'') or that the nuclear effects that MiniBooNE appears to
require vanish to a large extent at MINER$\nu$A's kinematics. With
regard to the last, work is in progress aimed at extending the
modeling of a relativistic 2p2h MEC analysis into the kinematical 
regime of MINER$\nu$A and NOMAD. Preliminary results indicate that 2p2h MEC 
effects might be expected to add about 12-15$\%$ 
to the IA results shown in this paper, in qualitative agreement with the findings of \cite{Nieves:2013min} and \cite{Mosel:2014lja}. What is reassuring at present is that 
the differences between the models at the higher-energy MINER$\nu$A kinematics 
are much smaller than for the MiniBooNE kinematics.
The good agreement between the IA predictions and MINER$\nu$A data resembles 
a similar situation for MiniBooNE data at forward scattering angles. 
On the contrary, this agreement gets lost for larger angles, 
which bear less weight at MINER$\nu$A kinematics.

Two additional issues have been addressed in the present study: one involves 
the use of kinematic cuts to elucidate the main contributions to the cross 
section (and showing how the high-energy MINER$\nu$A measurements are actually 
dominated by relatively small values of $q$ and $\omega$), while the other 
shows how Pauli blocking can be incorporated 
in the SuSA approach, improving the agreement at small values of $Q^2_{QE}$. 

\vspace{0.25cm}

The authors thank Laura Fields for her helpful comments and explanations on the experimental analysis. 
This work was partially supported by DGI
(Spain): FIS2011-28738-C02-01, FIS2011-24149, FPA2010-17142, by the
Junta de Andaluc\'ia (FQM-170, 225), by the INFN National Project
MANYBODY, the Spanish Consolider-Ingenio 2000 programmed CPAN, and
in part (T.W.D.) by US Department of Energy under cooperative agreement
DE-FC02-94ER40818, as well as by the Bulgarian National Science Fund
under contract DID-02/16-17.12.2009. G.D.M. aknowledges support from
a fellowship from the Fundaci\'on C\'amara (University of Sevilla).
M.V.I. is grateful for the warm hospitality given by the UCM and for
financial support during his stay there from the SiNuRSE action
within the ENSAR european project.

\end{document}